\documentclass[prb,reprint,amsmath,amssymb,floatfix,citeautoscript,superscriptaddress,longbibliography,noeprint]{revtex4-2}
\usepackage{cancel}
\usepackage{amsmath}
\usepackage{physics}
\usepackage{graphicx}
\usepackage{ulem}
\usepackage{amssymb}
\usepackage{sidecap}
\usepackage{amsthm}
\usepackage{bm}
\usepackage{dcolumn}
\usepackage{braket}
\usepackage{longtable}

\usepackage{ragged2e}
\usepackage{txfonts}
\usepackage{enumitem}
\setlist[enumerate]{label*=\arabic*.}

\allowdisplaybreaks

\usepackage{color}
\usepackage[usenames,dvipsnames]{xcolor}
\definecolor{myblue}{rgb}{0,0,1}
\usepackage[breaklinks=true,colorlinks=true,linkcolor=myblue,urlcolor=myblue,citecol  or=myblue]{hyperref}

\newcommand{\eps}{{\varepsilon}}

\begin{document}
\title{Conductivity of an electron coupled to anharmonic phonons}
\author{Jonathan H. Fetherolf}
\altaffiliation{Present address: Department of Chemistry, Yale University, New Haven, Connecticut 06520, USA}
\affiliation{Department of Chemistry, 
Columbia University, New York, New York 10027, USA}
\author{Petra Shih}
\affiliation{Department of Chemistry, 
Columbia University, New York, New York 10027, USA}
\author{Timothy C. Berkelbach}
\email{tim.berkelbach@gmail.com}
\affiliation{Department of Chemistry, 
Columbia University, New York, New York 10027, USA}
\affiliation{Center for Computational Quantum Physics, Flatiron Institute, New York, New York 10010, USA}

\begin{abstract}
We study the impact of phonon anharmonicity on the electronic dynamics of
soft materials using a nonperturbative quantum-classical approach. The method is applied
to a one-dimensional model of doped organic semiconductors with low-frequency intermolecular lattice phonons. We find that
anharmonicity that leads to phonon hardening increases the mobility and
anharmonicity that leads to phonon softening decreases the mobility. We also
test various approximations, including the use of adiabatic phonon disorder,
an effective harmonic model with temperature-dependent frequencies, and
the Boltzmann transport equation with second-order perturbation theory
scattering rates.  Overall, we find surprisingly good agreement between all
methods but that accounting for phonon anharmonicity is important for accurate
prediction of electronic transport including both quantitative mobility values
and their qualitative temperature dependence.  For the model studied, phonon lifetime effects have relatively little impact on carrier transport, but the effective frequency shift due to anharmonicity is essential.  In cases with highly asymmetric, non-Gaussian disorder, an effective harmonic model cannot quantitatively reproduce mobilities or finite-frequency conductivity, and this is especially true for acoustic phonons.

\end{abstract}

\maketitle
\section{Introduction}
Design of efficient functional materials requires a detailed microscopic
understanding of the charge transport mechanism.  In recent years, low-frequency
dynamic disorder has been shown to have a dominant role in the carrier dynamics
of soft semiconductors such as organic molecular crystals and lead halide
perovskites \cite{Troisi2006a,Fratini2015a,Mayers2018,Fetherolf2020}.  Like most
theories of electron-phonon interactions, the microscopic theory of dynamic
disorder commonly assumes linear coupling of carriers to harmonic phonons;
however, experimental and theoretical work suggests that the low-frequency
phonon modes in many materials---especially organic molecular crystals---exhibit
significant anharmonicity \cite{Asher2020,Schweicher2019,
Ruggiero2016,Ruggiero2017,Ferreira2020,Yaffe2017a,Alvertis2022}.  

In solids, anharmonic effects are responsible for thermal expansion, thermal
transport and structural phase transitions among other important nuclear effects
\cite{AshcroftMermin,Scott1974}.  With regards to spectral quantities, such as
vibrational spectroscopies, anharmonicity most commonly manifests as two
well-known and temperature-dependent effects: A frequency shift away from the
harmonic value (\textit{softening} or \textit{hardening}) and a finite
vibrational lifetime that introduces a spectral linewidth.  Unlike that of
harmonic phonons, the temperature dependence of anharmonic phonons may influence
the temperature dependence of coupled degrees of freedom, such as electrons or
excitons.  For example, a structural phase transition associated with anharmonic
mode-coupling has been shown to induce an insulator-metal transition in cuprates
\cite{Subedi2014,Mankowsky2016}.  Anharmonic modes are implicated in the
band-gap renormalization of several materials such as organic molecular crystals \cite{Alvertis2022}, halide perovskites
\cite{Patrick2015,Mayers2018} and strontium titanite \cite{Wu2020}.  Soft modes
in strontium titanite were also shown to be responsible for the specific
temperature dependence of the carrier mobility \cite{Zhou2018}, in a study that
made use of the perturbative Boltzmann transport equation with a
temperature-dependent effective phonons.  However, many soft materials have
strong electron-phonon coupling, which may preclude the use of perturbative
methods.  The effect of vibrational anharmonicity on electronic transport in
this context is not systematically known.  Previous theoretical studies on
organic molecular crystals have demonstrated that the details of harmonic
phonons, such as dispersion and symmetry, will lead to qualitative differences
in the carrier dynamics, particularly for temperature-dependent mobilities
\cite{Li2012,Li2013,Tu2018}.  In this paper, we use a nonperturbative
quantum-classical approach to study the motion of charge carriers coupled to
phonons with anharmonicity.

\section{Theory}
\label{sec:theory}

\subsection{Model Hamiltonian}

We study a single electron interacting with anharmonic phonons on a
one-dimensional lattice with fixed lattice constant $a$, $N$ sites, and periodic boundary conditions.  Each
lattice site $n$ has a single electronic orbital with creation operator
$c_n^\dagger$ and a single nuclear degree of freedom with momentum $p_n$ and
displacement $u_n$. 
For the model considered here, the electronic bands and 
phonons are trivially defined by symmetry.
The Hamiltonian is 
$H = H_{\mathrm{el}} + H_{\mathrm{ph}} + H_{\mathrm{el-ph}}$ with
\begin{subequations}
\label{eq:ham}
\begin{align}
\label{eq:ham_el}
H_{\mathrm{el}} &= -\tau\sum_{n} c^{\dagger}_n c_{n+1} + \mathrm{H.c.}
    = \sum_k \eps_k c_k^\dagger c_k, \\
\label{eq:ham_ph}
\begin{split}
H_{\mathrm{ph}} &= \sum_{n} \frac{p_n^2}{2m} + \mathcal{V}(u_1,\ldots,u_N) \\
    &= \sum_q \left[ \frac{p_q^2}{2} + \frac{1}{2}\omega_q^2 u_q^2 \right]
    + \mathcal{V}^\mathrm{an}(u_{q_1},\ldots, u_{q_N}),
\end{split} \\
\begin{split}
\label{eq:ham_elph}
H_{\mathrm{el-ph}} &= G \sum_{n} \left( c_n^\dagger c_{n+1} + \mathrm{H.c.}\right)
    \left(u_{n+1} - u_n\right) \\
    &= \sum_{kq} G_{kq}  c^\dagger_{k+q} c_{k} u_{-q},
\end{split}
\end{align}
\end{subequations}
where $\varepsilon_k$ are electronic band energies, $\omega_q$ are phonon frequencies, 
$\mathcal{V}^\mathrm{an}$ is the anharmonic part of the potential energy surface,
and we have assumed a linear Peierls form of the electron-phonon coupling
\begin{equation}
G_{kq} = \frac{2iG}{N^{1/2}} \left\{ \sin(ka) - \sin[(k-q)a] \right\}.
\end{equation}

We will study one model of optical phonons and one model of acoustic phonons according to
the potentials
\begin{subequations}
\label{eq:potentials}
\begin{align}
\label{eq:optical_anharm}
\mathcal{V}_\mathrm{op}(u_1,\ldots,u_N) &= \sum_{n} V(u_n), \\
\label{eq:acoustic_anharm}
\mathcal{V}_\mathrm{ac}(u_1,\ldots,u_N) &= \sum_{n} V(u_{n+1}-u_n), \\
V(u) &= \frac{1}{2} K u^2 + c_3u^3+c_4u^4.
\end{align}
\end{subequations}
Note that for simplicity we use the same function $V(u)$ in both cases.  In the
strictly harmonic limit, the above potentials yield optical phonons that are
dispersionless with $\omega_q = \omega_0$ and acoustic phonons with dispersion
$\omega_q = 2\omega_0|\sin(qa/2)|$, where $\omega_0 = \sqrt{K/m}$.
Numerical values of the parameters used in our simulations are given in Sec.~\ref{sec:results}.

\subsection{Nonperturbative and anharmonic simulation}

To accurately simulate the coupled electron-nuclear dynamics, we appeal to
the quantum-classical Ehrenfest approach~\cite{Troisi2006a,Wang2011a}.  Specifically, we let the nuclear degrees
of freedom evolve according to Newtonian dynamics on the anharmonic potential energy surface, 
$m \ddot{u}_n = -\partial \mathcal{V}/\partial u_n$.
From the nuclear trajectories $u_n(t)$, we define a time-dependent electronic Hamiltonian
\begin{equation}
\begin{split}
h_\mathrm{el}(t) &= -\sum_n \tau_n(t) 
        \left(c_n^\dagger c_{n+1} + \mathrm{H.c.} \right) \\
    &= \sum_{kq} \left[ \eps_k \delta_{k+q,k} + G_{kq} u_{-q}(t) \right]
        c_{k+q}^\dagger c_k
\end{split}
\end{equation}
where $\tau_n(t) = \tau - G[u_{n+1}(t)-u_n(t)]$.
Following our previous work~\cite{Fetherolf2020}, 
this time-dependent Hamiltonian is used in a mixed quantum-classical evaluation of the electronic current 
autocorrelation function,
\begin{equation}
C_{jj}(t) = \int d\bm{p}\int d\bm{u} \ \mathcal{P}(\bm{p},\bm{u})
    \langle U_{\mathrm{el}}(0,t) j(t) U_{\mathrm{el}}(t,0) j \rangle_{\mathrm{el}}.
\end{equation}
Here, $\mathcal{P}(\bm{p},\bm{u}) \sim e^{-H_\mathrm{ph}(\bm{p},\bm{u})/k_\mathrm{B}T}$ is the phase-space distribution 
of the classical nuclear degrees of freedom, 
\begin{equation}
\label{eq:dd_evol}
U_\mathrm{el}(t,0) = T \exp\left[-\frac{i}{\hbar} \int_0^t dt^\prime 
        h_\mathrm{el}(t^\prime) \right]
\end{equation}
is the time-ordered evolution operator,
\begin{equation}
\label{eq:current_op}
j(t) = ia\sum_n \tau_n(t) 
        \left(c_n^\dagger c_{n+1} - \mathrm{H.c.} \right)
\end{equation}
is the current operator for the Hamiltonian~(\ref{eq:ham}),
$\langle O \rangle_{\mathrm{el}} = \mathrm{Tr}_{\mathrm{el}}\{Oe^{-h_{\mathrm{el}}(t=0)/k_\mathrm{B}T}\}/Z_{\mathrm{el}}$ is
a thermal average over electronic degrees of freedom, and 
$Z_\mathrm{el} = \mathrm{Tr}_{\mathrm{el}}\{e^{-h_\mathrm{el}(t=0)/k_\mathrm{B}T}\}$ is the
electronic partition function.
From this, the AC conductivity is readily obtained via the Kubo formula
\begin{equation}
\mathrm{Re} \sigma(\omega) = \frac{1-e^{-\hbar\omega/k_\mathrm{B}T}}{2N\omega} 
    \int_{-\infty}^{+\infty} dt\ e^{i\omega t} C_{jj}(t),
    \label{eq:kubo}
\end{equation}
with the DC component obtained by taking the zero-frequency limit,
$\sigma_{\mathrm{DC}}\equiv \sigma(\omega\rightarrow 0)$.
The mobility is $\mu = (Na/e)\sigma_\mathrm{DC}$, where $e$ is the electron charge.
The nonperturbative and anharmonic approach described here will be referred to as the ``dynamical Kubo''
approach in later sections. 

Previous studies of conductivity performed via the Kubo formula have generally
been done in the adiabatic limit~\cite{Fratini2009,Cataudella2011,Li2012,Li2013} such that
Eq.~(\ref{eq:kubo}) reduces to 
\begin{equation}
\label{eq:statickubo}
\begin{split}
    \mathrm{Re} \sigma(\omega) &= \frac{1-e^{-\hbar\omega/k_\mathrm{B}T}}{2N\omega} 
    \int d\bm{u} \ \mathcal{P}(\bm{u}) \\
&\hspace{3em}\times Z_\mathrm{el}^{-1} \sum_{\alpha\beta}
    |\langle \alpha|j|\beta\rangle|^2\delta(\omega-(\eps_\beta-\eps_\alpha)/\hbar),
\end{split}
\end{equation}
where $\alpha,\beta$ are $\bm{u}$-dependent eigenstates of the disordered electronic
Hamiltonian and $\mathcal{P}(\bm{u})\sim e^{-\mathcal{V}(\bm{u})/k_\mathrm{B}T}$.
We will refer to this as the ``static Kubo'' approach.
One drawback is that the delta function must be given an artificial 
linewidth $\eta$, which is analogous to imposing an artificial
decay to the current autocorrelation function (also called 
the ``relaxation time approximation'' in the transient localization 
literature~\cite{Fratini2015a,Ciuchi2011}).
The choice of $\eta$ has been shown to substantially
affect the temperature dependence of the mobility~\cite{Cataudella2011};
this ambiguity is avoided in the dynamical Kubo approach,
where the current autocorrelation function decays naturally due to
dynamic disorder (for both harmonic and anharmonic phonons).

The quantum-classical Ehrenfest approach has been one of the preferred methods for
calculating carrier dynamics in models of soft 
materials~\cite{Troisi2006a,Wang2011a,Fratini2015a,Fetherolf2020}. Because of its
nonperturbative nature, this method can be applied to
materials with simultaneously large electronic transfer integral $\tau$ and
large electron-phonon coupling $G$, a combination which precludes treatment with
perturbative small polaron or Boltzmann transport theory~\cite{Holstein2000,Mahan2000}.
The classical treatment of the phonon degrees of freedom is
approximate, but highly accurate when $\hbar\omega_0/k_\mathrm{B}T$ is small, as is the case for many
soft semiconductors, including those we study here.  The accuracy of
quantum-classical methods was previously verified via comparison with accurate
quantum Monte Carlo results with analytic continuation~\cite{DeFilippis2015,Fratini2015a}. 
In addition to the classical approximation
for nuclear dynamics, we also neglect the feedback of the electronic degrees of
freedom on the nuclei; this approximation has been shown to be accurate for the
low-frequency nuclear dynamics and relatively high-mobility parameter regime 
we study here~\cite{Wang2011a}.

\subsection{Perturbation theory with harmonic phonons}

In fully ab initio studies, it is most common to calculate electronic dynamics by neglecting anharmonicity
and treating the electron-phonon interaction by perturbation 
theory~\cite{Verdi2015,Bernardi2016,Mustafa2016,Giustino2017,Liu2017}; see Ref.~\onlinecite{Lee2018} for application
to naphthalene, an organic molecular crystal similar to the model that we study in this work.
This approach yields the intraband scattering rates 
\begin{equation}
\label{eq:btept_harm}
\begin{split}
\Gamma_{k,k+q}(T) &= \frac{\pi}{\hbar\omega_q} |G_{kq}|^2
    \Big\{[n_q+1] \delta(\eps_k-\eps_{k+q}-\omega_q) \\
&\hspace{6em} 
    + n_q \delta(\eps_k-\eps_{k+q}+\omega_q)\Big\}
\end{split}
\end{equation}
and inverse lifetimes $\Gamma_k(T) = \sum_q \Gamma_{k,k+q}(T)$;
here, $n_q$ is the Bose-Einstein distribution function at temperature $T$.
In the limit of low $\omega_q$, which holds for the materials of interest here, 
the quasielastic approximation can be made~\cite{Mahan2000,Li2013} giving
\begin{equation}
\label{eq:bte_elastic}
\begin{split}
\Gamma_{k,k+q}(T) &= \frac{\pi \omega_0 k_\mathrm{B}T}{\hbar\omega_q} |G_{kq}|^2 
    \delta(\eps_k-\eps_{k+q})\Big\{1-\cos(\theta_{k,k+q})\Big\},
\end{split}
\end{equation}
where $\omega_0$ was defined below Eq.~(\ref{eq:potentials}), 
$\theta_{k,k+q}$ is the angle between the initial and the scattered state, and $1-\cos(\theta_{k,k+q})=1-(k+q)/k$ in one dimension.
These lifetimes can be used within a linearized Boltzmann transport equation
(BTE) framework~\cite{Bernardi2016,Hamaguchi2017} to calculate the conductivity,
\begin{equation}
\sigma_\mathrm{DC} = \frac{e^2}{Nak_\mathrm{B}T} Z_\mathrm{el}^{-1} \sum_k v_k^2 \Gamma_k^{-1} e^{-\eps_k/k_\mathrm{B}T}
\end{equation}
where $v_k=\hbar^{-1}\partial \eps_k/\partial k$ is the band velocity.  The
quasi-elastic BTE provides a useful comparison for the limit of electronic band
transport with weak scattering due to phonons.  As lowest-order perturbation theory,
the BTE ignores multi-phonon processes that mediate relaxation when the electronic energy difference
and the phonon energies are mismatched. Thus we expect the BTE to overestimate electronic lifetimes and
therefore overestimate the conductivity. These multiphonon processes are captured in the 
nonperturbative quantum-classical theory.

\subsection{Effective harmonic theory}

The phonon anharmonicity may be treated approximately by using an effective, temperature-dependent harmonic model,
\begin{equation}
\mathcal{V}(u_{q_1},\ldots,u_{q_n}) \approx \frac{1}{2} \sum_q \tilde{\omega}_q^2(T)\ u_q^2 
\end{equation}
where the effective phonon frequencies $\tilde{\omega}_q(T)$ can be determined
by a number of mean-field type methods~\cite{Hooton1958,Werthamer1970,Souvatzis2008,Hellman2013,Tadano2015}.
Specifically, we highlight Ref.~\onlinecite{Zhou2018}, which treats the dynamics of electrons coupled to 
soft modes in SrTiO$_3$ in this manner. 
Here we propose an alternative but closely related approach motivated by the
application to electronic dynamics.

For the anharmonic potentials considered here, 
we will consider effective harmonic potentials of the same form as in Eqs.~(\ref{eq:potentials})
but with
\begin{equation}
\tilde{V}(u) = \frac{1}{2}\tilde{K} u^2 = \frac{1}{2} m \tilde{\omega}_0^2 u^2,
\end{equation}
and we will employ the same form of electron-phonon coupling as in Eq.~(\ref{eq:ham_elph}).
The temperature-dependent effective frequency $\tilde{\omega}_0$ is chosen to reproduce the statistics of 
the dynamically disordered transfer integrals $\tau_n(t)$.
Our later simulations will be performed at fixed volume (i.e., in the absence of thermal expansion), which
guarantees that the average transfer integral is always given by the bare transfer integral, 
$\langle \tau_n \rangle = \tau$, even with anharmonicity.
We thus choose the effective harmonic frequency to reproduce the variance of the transfer integral calculated with
the anharmonic potential, $\langle (\tau_n-\tau)^2 \rangle_\mathrm{ha} = \langle (\tau_n-\tau)^2 \rangle_\mathrm{an}$, 
which is equivalent to matching the variance in the nearest-neighbor separations,
$\langle (u_{n+1}-u_n)^2 \rangle_\mathrm{ha} = \langle (u_{n+1}-u_n)^2 \rangle_\mathrm{an}$. This requirement
leads to
\begin{equation}
\tilde{\omega}_0^2 = C \frac{2G^2k_\mathrm{B}T}{\langle (\tau_n-\tau)^2\rangle_\mathrm{an}}
    = C \frac{2k_\mathrm{B}T}{\langle (u_{n+1}-u_n)^2\rangle_\mathrm{an}},
\label{eq:effectivefreq}
\end{equation}
where $C=2$ for optical phonons and $C=1$ for acoustic phonons. We emphasize that the anharmonic variance 
is a statistical quantity 
that can be calculated with Monte Carlo sampling and does not require any information
about the dynamics of the phonons.  
This formalism is closely related to other mean-field theories of phonon anharmonicity.

This approach approximately captures the instantaneous electronic disorder in
the Hamiltonian, which is the primary ingredient of the transient
localization/dynamic disorder picture~\cite{Fratini2015a,Troisi2006b}. In the
next section, simulations using this effective harmonic model of phonons (with
the dynamical Kubo, static Kubo, and Boltzmann transport theories described
above) will be compared to fully anharmonic simulations. In this way, we can
isolate the effects of anharmonicity (treated approximately or exactly) and
nonperturbative electron-phonon coupling.

\section{Results and Discussion}
\label{sec:results}
\subsection{Simulation details and phonon anharmonicity}
As a harmonic limit for our transport model, we use the parameters from 
Ref.~\onlinecite{Troisi2007} for the $b$-axis of single-crystal rubrene, 
which have been used in numerous studies~\cite{Cataudella2011,Fratini2015a,Fetherolf2020}.  The parameters are
$\tau=143$~meV, $G=493.5$~meV/\AA, 
$m=532$~amu and $a=7.2$~\AA.  
As already mentioned, we do not allow thermal expansion, which guarantees $\langle\tau_n\rangle=\tau$.
We used periodic lattices with 100-200 sites and sampled up to 50~000 trajectories for each calculation to
converge all results.

For simplicity, we consider the same set of potential parameters $K,c_3,c_4$ for the
optical and acoustic phonons. 
The harmonic force constant is always $K=m\omega_0^2=4.89$~eV/\AA$^2$, corresponding
to $\hbar\omega_0 = 6.2$~meV.
For each phonon model, we consider two type of anharmonicity.
For the first type, which we call ``phonon hardening'', we use $c_3=0$ and $c_4 = 19.56$~eV/\AA$^4$;
this type of purely quartic anharmonicity will result in an increase in the phonon
frequency with temperature.
For the second type, which we call ``phonon softening'', we use $c_4 = 2.45$~eV/\AA$^4$
and two possible values for $c_3$. The first value, $c_3=-4.40$~eV/\AA$^3$,
yields an asymmetric single-well potential; the second value, $c_3=-4.65$~eV/\AA$^3$,
yields an asymmetric double-well potential.

\begin{figure}
    \centering
    \includegraphics{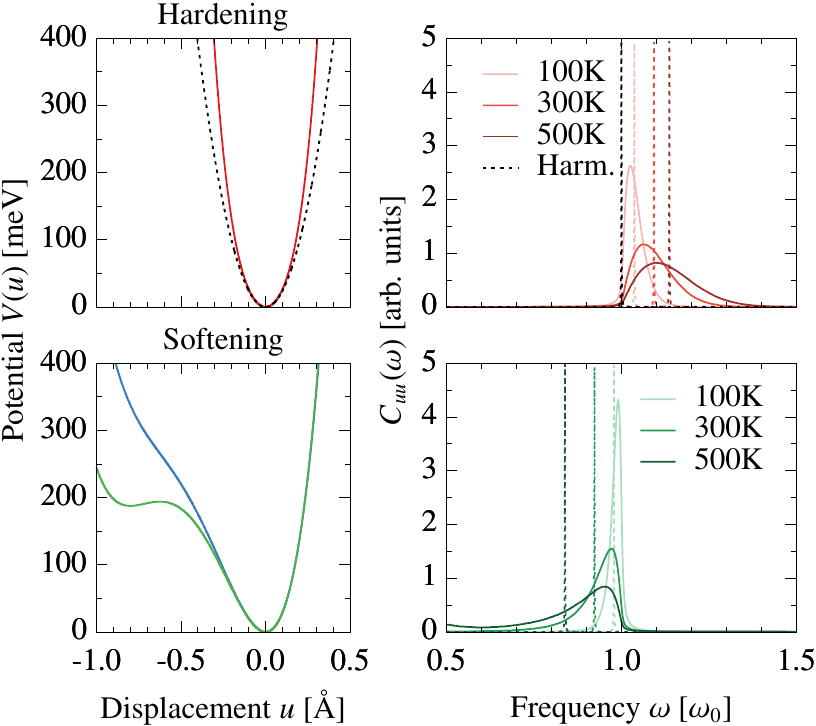}
    \caption{
Potential $V(u)$ (left) and spectral function $C_{uu}(\omega)$ (right)
for optical phonons.  The phonon potential is of the form in Eq.
\ref{eq:optical_anharm} with hardening (top left, red) corresponding to $c_3=0$
and $c_4=19.56$ eV/\AA$^4$.  The phonon softening potentials (bottom left) both use
$c_4=2.45$ eV/\AA$^4$ with $c_3=-4.40$ and $-4.65$ eV/\AA$^3$ for the single
and double well, respectively.  The dashed black curve is the harmonic potential
with $c_3=c_4=0$ (top left).  The spectral function $C_{uu}(\omega)$ is shown at
different temperatures using the hardening parameters (top right) and the
double-well softening parameters (bottom right).  The dashed lines correspond to
the effective harmonic frequencies $\tilde{\omega}_0(T)$ obtained using Eq.
\ref{eq:effectivefreq}.}
    \label{fig:potentials}
\end{figure}

In Fig.~\ref{fig:potentials}, we plot these three potentials (left column) and
the temperature-dependent spectral function $C_{uu}(\omega)\sim \mathrm{Re}\int dt e^{i\omega t}
\langle u(t)u(0) \rangle_\mathrm{an}$ (right column) for a single anharmonic
oscillator calculated with classical dynamics. Unlike for a harmonic potential, the anharmonic potentials lead to
spectral functions whose peaks shift to higher frequencies (phonon hardening) or lower
frequencies (phonon softening) with increasing temperature, along with a decrease
in the phonon lifetimes. We also
plot the temperature-dependent effective frequencies (vertical dashed lines)
determined by matching the variance $\langle u^2\rangle$ between the anharmonic and
effective harmonic potential. The variance is related to the spectral function
by $\langle u^2\rangle = \int d\omega C_{uu}(\omega)$ and therefore the effective harmonic
frequency is determined by $\tilde{\omega}_0^2 = k_\mathrm{B}T / \int d\omega C_{uu}(\omega)$.
We note that while the effective frequency roughly matches the \textit{maximum} of the spectral function
in the model with phonon hardening, it deviates more strongly from the maximum 
in the model with phonon softening. This is due to the more asymmetric distribution of the latter,
whose \textit{maximum} does not shift significantly with temperature despite the development of
a large tail extending to lower frequencies.

\begin{figure}[t]
    \centering
    \includegraphics{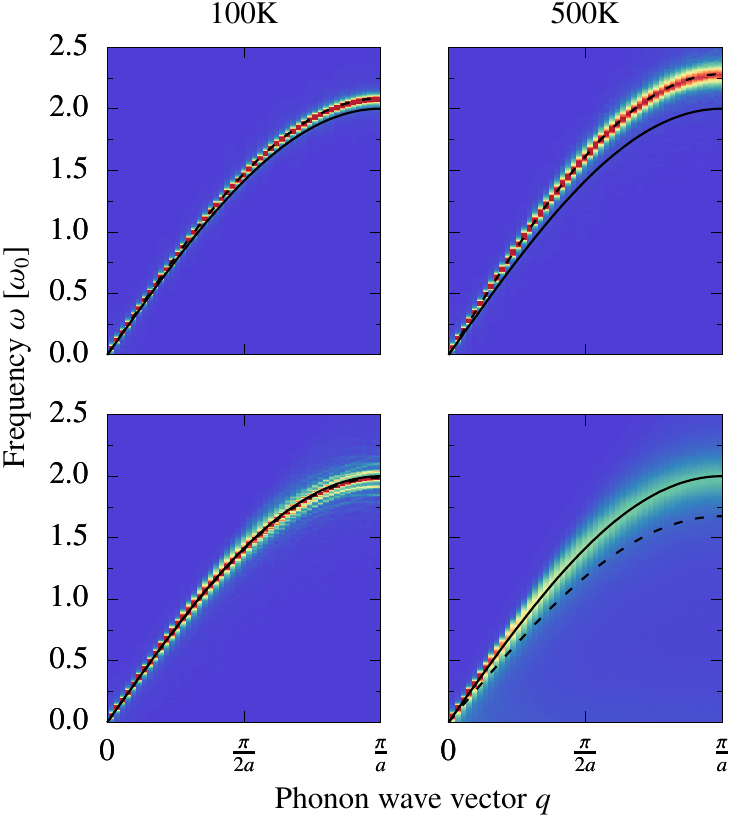}
    \caption{Momentum-resolved phonon spectral function $C_{uu}(q,\omega)$ for
acoustic phonons.  The color map data show the spectra for the hard mode
potential (top) and double-well soft mode potential (bottom) at two different
temperatures.  The solid black line shows the harmonic dispersion
$2\omega_0|\sin(qa/2)|$, while the dashed black line shows the effective
harmonic dispersion $2\tilde{\omega}_0(T)|\sin(qa/2)|$.}
    \label{fig:Ckw_acoustic}
\end{figure}

When these potentials are used as the pair potential for our model of acoustic phonons,
they yield the momentum resolved phonon spectral function 
$C_{uu}(q,\omega) \sim \mathrm{Re} \int dt e^{i\omega t} \langle u_q(t) u_{-q}(0)\rangle$
shown in Fig.~\ref{fig:Ckw_acoustic}; results are shown at two temperatures.
With quartic anharmonicity, we see clear phonon hardening of the entire
phonon branch, with a peak position that is well-matched by the effective harmonic
dispersion $\tilde{\omega}_q = 2\tilde{\omega}_0|\sin(qa/2)|$, and a decreased phonon lifetime.
For the phonon softening case, there is a clear decrease in the phonon lifetime. The effective phonon
frequency shows the expected signature of phonon softening, although the spectral structure
of the anharmonic result is hard to see with the employed colorscale due to the large linewidth. Like in Fig.~\ref{fig:potentials},
the maximum of the spectral function does not shift significantly but a large tail develops that
extends to low frequency, which is captured in an average sense by the effective harmonic model.

The phonon model parameters were chosen to represent physically realistic amounts of
anharmonicity.  In particular, the frequency shifts and broadenings observed in
Figs.~\ref{fig:potentials} and \ref{fig:Ckw_acoustic} are comparable to those observed in low-frequency
Raman measurements and ab initio simulations of organic crystals
\cite{Asher2020,Schweicher2019}.

\subsection{Optical phonons}
\begin{figure}
    \centering
    \includegraphics{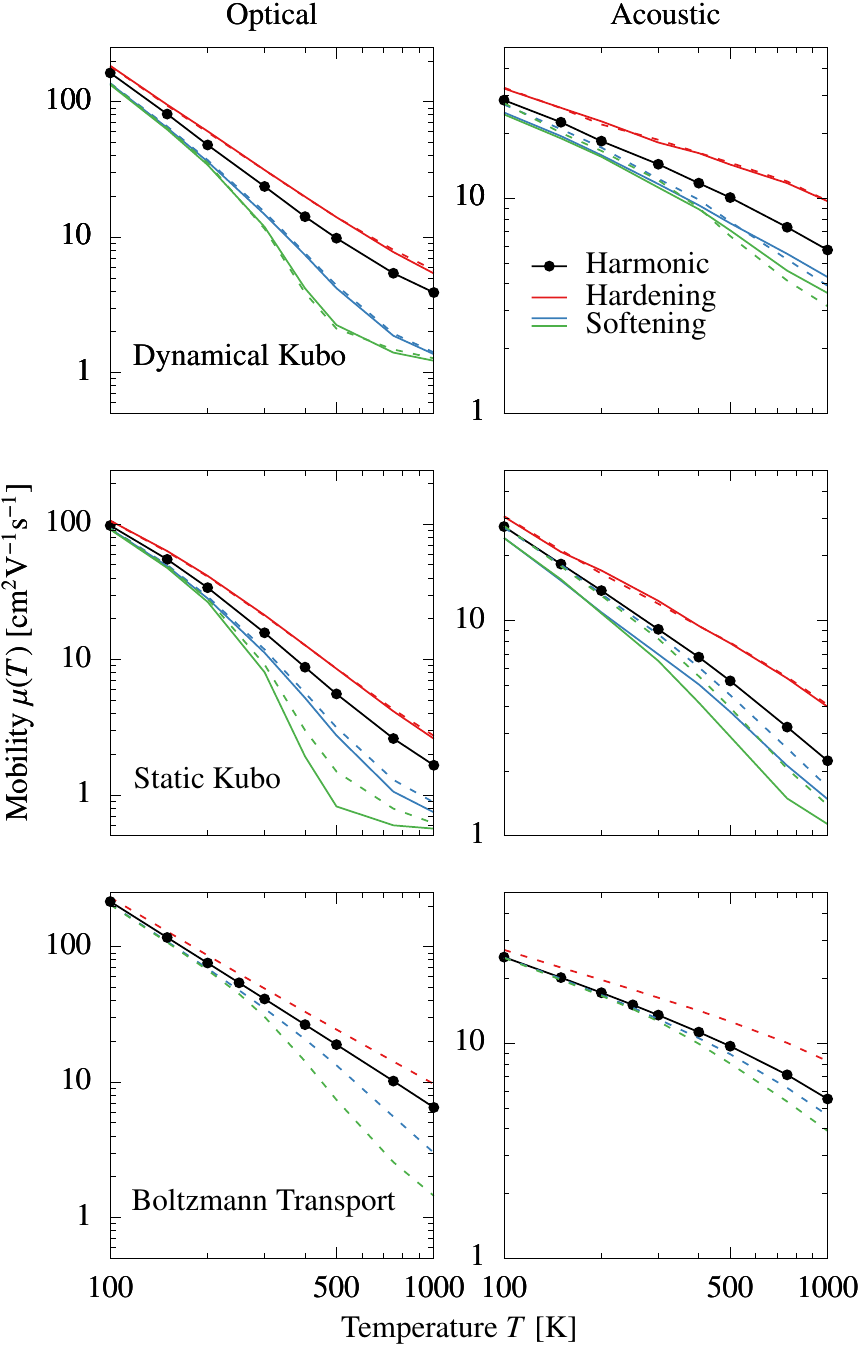}
    \caption{Log-log plot of temperature-dependent mobility $\mu(T)$ for a
carrier coupled to optical phonons (left) or acoustic phonons (right).
Mobility is calculated with the dynamical Kubo formula (top), the static Kubo
formula with $\eta=\omega_0/2$ (middle), and the Boltzmann transport
equation (bottom).  In addition to the mobility with harmonic phonons (black
dotted lines), we show mobility with anharmonic phonons that lead to hardening
(solid red) and softening (blue and green).  Results are shown with full
anharmonicity (solid) and with temperature-dependent
harmonic phonons (dashed) according to Eq.~(\ref{eq:effectivefreq}).}
    \label{fig:mobility}
\end{figure}
Having constructed a model for phonon anharmonicity in solids, we now shift our attention to the impact on electronic transport.
On the left hand side of Fig.~\ref{fig:mobility} we present the mobility of a
carrier coupled to optical phonons, using the three levels of theory described in
Sec.~\ref{sec:theory}:
the dynamical Kubo approach, the static Kubo approach, and the Boltzmann
transport equation (BTE).
We first consider the harmonic limit with $c_3=c_4=0$ (black circles).  
The dynamical Kubo approach yields a power law of roughly $\mu \propto T^{-1.8}$,
consistent with other Kubo formula
calculations \cite{Cataudella2011,Li2012,Li2013,Fratini2015a} and
Ehrenfest-style mixed quantum-classical diffusion models
\cite{Troisi2006a,Troisi2007,Troisi2011,Troisi2011b}.  We see that the
``band-like'' power law behavior extends to low temperatures, but the mobility
begins to saturate above 500~K.
This high-temperature mobility saturation is a well-known result of the
nonperturbative quantum-classical models, equivalent to resistivity saturation
in metals~\cite{Millis1999,Fratini2009}.  The static Kubo approach with
$\eta=\omega_0/2$ produces similar features but exhibits different
low-temperature behavior and a different power law exponent.  Both of these
differences between static and dynamical Kubo formula are due to the artificial
lifetime $\eta^{-1}$ in the static Kubo formula.  The dynamical Kubo formula results
in a natural lifetime that changes with temperature, most notably at low
temperature.  Finally, the quasielastic BTE predicts a power law of $T^{-1.5}$
at low-temperature and $T^{-2}$ at
high-temperature~\cite{Fratini2009,Li2012,Li2013}; only the former is visually
apparent for the current model parameters within the temperature range shown.  
Despite the differences in their detailed behaviors, all methods yield absolute
mobilities that are within a factor of two of the ``exact'' dynamical Kubo approach
at all temperatures.

We now consider the impact of phonon anharmonicity. 
Qualitatively, we expect phonon hardening to increase the mobility and phonon softening
to decrease the mobility, because 
electrons near the bottom of the band are more effectively scattered by low-frequency
phonons.
Indeed, in our simulations, we see that purely quartic anharmonicity leading
to phonon hardening (red lines)
increases the overall mobility (by no more than 50\%) and slightly decreases the
power law coefficient. 
Both types of anharmonicity leading to phonon softening,
an asymmetric single well (blue lines) and a double well (green lines),
reduce the mobility---by up to a factor of three
in the high temperature limit of the dynamical Kubo results.
Moreover, phonon softening lowers the onset of ``high-temperature'' behavior;
all methods show a crossover to $\mu \sim T^{-2}$ power law behavior but only
the nonperturbative Kubo results (dynamical or static) show the later onset 
of mobility saturation~\cite{Fratini2009}.
For the double-well potential, the anharmonic frequency shift is larger and thus
the reduction in the mobility is larger.
Compared to the exact dynamical Kubo method, the static Kubo method overestimates 
the reduction in the mobility.

Replacing the anharmonic phonons by effective harmonic phonons (dashed lines) is seen to be
an excellent approximation. For the dynamical Kubo approach, the agreement is almost perfect, 
suggesting that the phonon lifetime has no appreciable effect on the mobility.
This is not surprising given the separation of timescales for the parameters used here:
the electronic lifetimes are hundreds of femtoseconds and the phonon lifetimes are 
thousands of femtoseconds, even at high temperature.  
Within the static Kubo framework, discrepancies between mobilities with the
effective harmonic and with fully anharmonic potential are due to the
shape of the disorder distribution.  While both cases have the same
electronic disorder variance $\langle (\tau_n-\tau)^2 \rangle$, the
fully anharmonic potential produces a highly non-Gaussian distribution,
which modifies the electronic dynamics.  
The agreement is best for purely quartic (hardening) anharmonicity because the
distribution is symmetric and thus can be accurately modeled by an appropriate
Gaussian distribution; the asymmetric distributions associated with phonon softening
present a greater challenge.
Another contributor to the static Kubo 
mobility is the electronic lifetime $\eta^{-1}$.  We used a constant $\eta=\omega_0/2$ regardless of the
effective frequency $\tilde{\omega}_0(T)$ for the static Kubo results; using
$\eta=\tilde{\omega}_0(T)/2$ was found to produce an even larger discrepancy with the
dynamical Kubo results and a large overestimate of the effect of phonon
softening on carrier mobility.

The BTE framework qualitatively captures the effect of phonon hardening; in fact,
the mobility is just scaled by a factor of $\tilde{\omega}_0(T)/\omega_0$.
Phonon softening accentuates the
disagreement between BTE and the nonperturbative Kubo results, since it
reduces the onset temperature of mobility saturation, which cannot be captured by the BTE.
In summary, the DC mobility of an electron coupled to anharmonic optical phonons is
accurately reproduced by effective harmonic phonons within the full
dynamical Kubo formula; this approximation becomes less accurate for the static Kubo
approach. This effective harmonic approximation is the only way to apply BTE,
which becomes increasingly inaccurate at high temperature, especially in the presence of phonon
softening.

\begin{figure}
    \centering
    \includegraphics{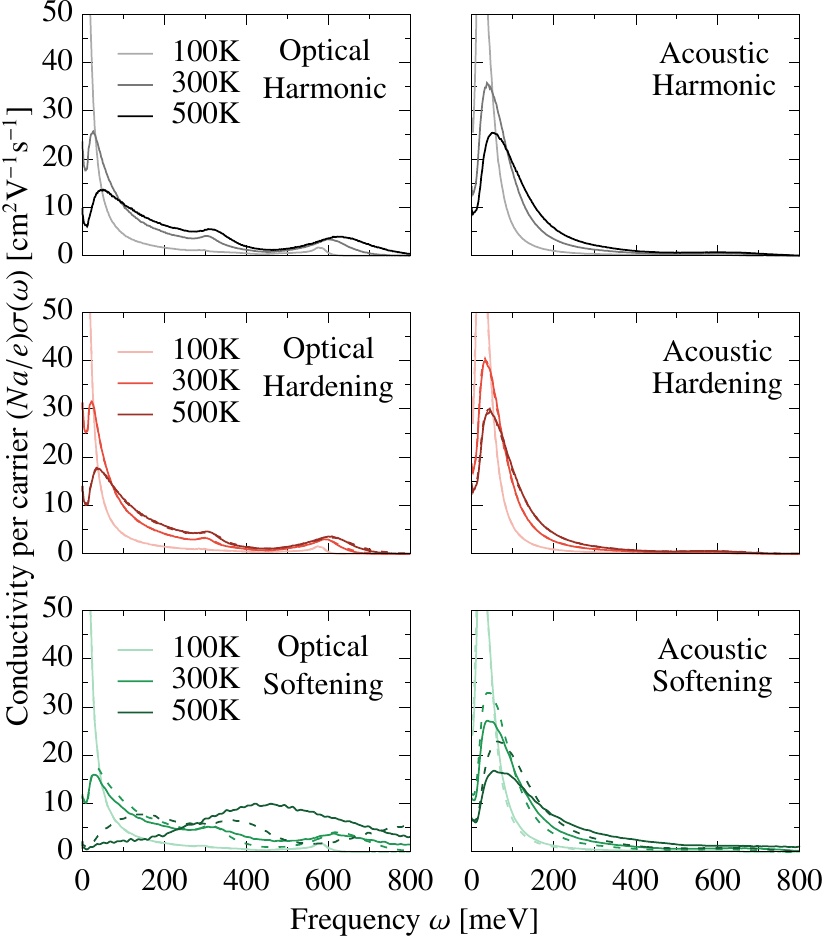}
    \caption{AC conductivity per carrier $(Na/e)\sigma(\omega)$ with coupling to
optical (left) and acoustic phonons (right).  Conductivities are calculated
using the dynamical Kubo formula (solid lines) and are shown at different temperatures for
harmonic phonons (top), anharmonic phonons that harden with temperature
(middle), and double-well anharmonic phonons that soften with temperature
(bottom). Dashed lines use an effective harmonic potential with $\tilde{\omega}_0$
from Eq.~(\ref{eq:effectivefreq}).
}
    \label{fig:sigma}
\end{figure}

In Fig.~\ref{fig:sigma}, we show the frequency-resolved AC conductivity of the same systems
calculated with the dynamical Kubo formula at three temperatures: 100~K, 300~K, and 500~K.
In the harmonic limit, we see the characteristic features of asymmetric
electron-phonon coupling to optical phonons~\cite{Cataudella2011,Li2011,Li2012}.
Peaks appear at low frequency (below 50 meV) and at multiples of the half
bandwidth $2\tau$ (around 300 and 600 meV).  
With increasing temperature, the spectral weight shifts to higher energies
and the DC mobility is reduced.
In the presence of pure quartic anharmonicity (phonon hardening), we see very similar
results, although the anharmonicity induces a slight shift of spectral weight to lower energies,
consistent with the increased mobility. The effective harmonic model produces
nearly exact results at all frequencies, extending the agreement seen in the mobility.
When the anharmonicity is of the double-well form (phonon softening),
we find a low-temperature conductivity similar to the harmonic or quartic anharmonic cases, 
but quite different high-temperature conductivities with significantly less structure,
While the effective harmonic model can reproduce the DC conductivity of the anharmonic model
reasonably well, it is less successful for the full AC conductivity, especially at higher temperatures.
Results obtained with the static Kubo formula (not shown) confirm that this discrepancy
is not a phonon lifetime effect, but rather is due to the harmonic
approximation's inability to capture the non-Gaussian distribution of electronic
disorder.



\subsection{Acoustic phonons}


We now study an electron coupled to anharmonic acoustic phonons, recalling
that it has been previously demonstrated
that, \textit{within the harmonic approximation}, acoustic phonons and optical phonons
lead to qualitatively different transport behavior~\cite{Li2013,Tu2018}. 
The right-hand side of Fig.~\ref{fig:mobility} shows the temperature
dependent mobility. For harmonic acoustic phonons, the dynamical Kubo formula
predicts a power law of roughly $\mu \propto T^{-1/2}$ below 500~K.  In
the regime where saturation occurs for optical phonons, the mobility has the
opposite behavior for acoustic phonons, showing an increased power law
coefficient; mobility saturation will occur at very high temperature 
but is not evident for harmonic acoustic phonons at the temperatures 
shown \cite{Li2013}.  Static Kubo and BTE calculations show similar behavior, 
although the power law exponent of the static Kubo mobility is overestimated.
Surprisingly, the BTE mobility is more accurate than the static Kubo mobility,
presumably due to the latter's use of a constant $\eta$.
In agreement with the dynamical Kubo results, the BTE predicts $\mu \propto T^{-1/2}$ at 
low temperature at $\mu \propto T^{-2}$ at high temperature.  

Turning to the impact of anharmonicity, we see that acoustic phonon hardening 
produces the expected behavior based on our previous
analysis: it increases the mobility and reduces
the power-law coefficient, effects that are qualitatively captured by all three
methods. Acoustic phonon softening decreases the mobility and introduces mobility
saturation at lower temperatures. 
Unlike for optical phonon anharmonicity, phonon hardening generally 
modifies the mobility more than phonon softening.


Again we find that phonon hardening effects are well reproduced by an effective harmonic
model. However, phonon softening is even harder to capture with a harmonic model 
than it was for optical phonons; the mobility is overestimated at all temperatures
and disagreement is most severe for the static Kubo approach.
The BTE mobility is a remarkably good approximation to the dynamical Kubo results
and semiquantitatively captures the effects of all types of anharmonicity.


The frequency-resolved AC conductivity calculated by the dynamical Kubo approach
with harmonic and anharmonic acoustic phonons is shown in the right hand side of
Fig.~\ref{fig:sigma}.
With coupling to acoustic phonons, the AC conductivity is significantly 
different than with coupling to optical phonons.
Specifically, there is absolutely no structure at high frequencies; only a
DC conductivity and a simple maximum at low frequency.
For this reason, the effective harmonic approximation is more successful than
it was or optical phonons. The agreement is worst with phonon softening, where
the effective harmonic approximation slightly underestimates the linewidth.

\begin{figure}
    \centering
    \includegraphics{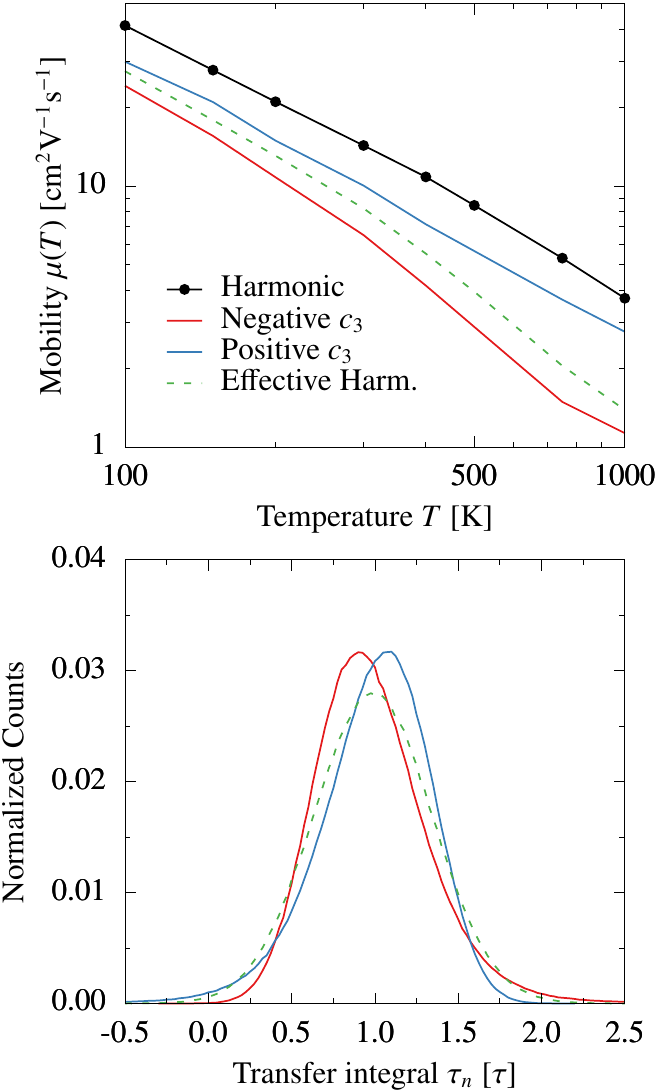}
    \caption{Static Kubo formula mobility with $\eta=\omega_0/2$ (top
panel) and example distributions of transfer integral $\tau_n$ at $500$~K (bottom
panel) for an electron coupled to a soft acoustic mode. All distributions in the
bottom panel share the same average transfer integral $\langle \tau_n \rangle=\tau$ and
variance $\langle (\tau_n-\tau)^2 \rangle=0.123\tau^2$, with the red and blue curves
differing only in the \textit{sign} of the cubic coefficient $c_3=\pm 4.65$
eV/\AA$^3$.  Both potentials lead to the same effective harmonic distribution of
$\tau_n$ and mobility (green dashed lines).
}
    \label{fig:dist_acoustic}
\end{figure}

Before concluding, we give an example of the challenge associated with effective
harmonic approximations.
We consider two version of the asymmetric double well pair potential, leading 
to acoustic phonon softening: one with $c_3 = -4.65$ eV/\AA$^3$ (the same
one considered so far) and one with $c_3 = +4.65$ eV/\AA$^3$.
In the bottom panel of Fig.~\ref{fig:dist_acoustic}, we show the distibution
of nearest-neighbor transfer integrals $P(\tau_n)$. Both anharmonic potentials have the same
mean $\langle \tau_n \rangle$ and variance $\langle (\tau_n-\tau)^2 \rangle$; therefore,
the effective harmonic potential constructed according to our
prescription~(\ref{eq:effectivefreq}) is identical.
The top panel of Fig.~\ref{fig:dist_acoustic} shows the static Kubo mobility 
for each of these three potentials. We see that the
magnitude and temperature dependence of the mobility differs depending on the
sign of $c_3$, an effect that cannot be captured by an effective harmonic model.  
While asymmetric potentials are difficult to capture using a harmonic model for
optical modes, the issue of asymmetric distributions of $\tau_n$ is not present
in (strictly dispersionless) optical phonons.
This is because $u_{n+1}$ and $u_n$ are uncorrelated, which yields a
symmetric distribution for $\tau_n$ even for asymmetric potentials.  Correlated
nearest-neighbor displacements, like those in acoustic phonons, produce an 
asymmetric distribution of $\tau_n$ and thus create the ambiguity demonstrated here. 
%

\section{Conclusions}
\label{sec:conclusions}


We have demonstrated the impact of phonon anharmonicity on the equilibrium
electronic dynamics of soft materials and assessed the accuracy of various approximations.
For all methods,
we see a change in the magnitude and temperature dependence of the mobility based on
the strength and form of the anharmonicity.
Within the dynamical Kubo formula, changes to the mobility are
well-characterized by an effective harmonic model of phonons, especially for optical
phonons.  
The effective harmonic model is less accurate when the static Kubo formula is employed.
This discrepancy is especially apparent for acoustic phonons, where the correlated phonon motion
can lead to asymmetric disorder profiles that cannot be
unambiguously modeled by a harmonic potential. 

Future work could include lattice expansion or lattice strain, studies of which
have been mostly limited to harmonic models~\cite{Ruggiero2019,Landi2021}.
The approaches described are quite general and could be applied to other soft
materials such as metal-oxide perovskites~\cite{Wu2020}, lead-halide 
perovskites~\cite{Patrick2015,Mayers2018}, or thermoelectric materials~\cite{Zhao2020,Hu2021},
perhaps in an ab initio framework. We are also interested in applying these methods
to study nonequilibrium electronic dynamics~\cite{Bernardi2014,Jhalani2017} where we expect the impact of
nonperturbative electron-phonon interactions and phonon anharmonicity to be larger.


\section*{Acknowledgements}
We thank Prof.~Omer Yaffe for helpful discussions.  This work was supported in part
by the NSF under Grant No.~DGE-1644869 and by the NSF MRSEC program through Columbia in the Center for
Precision Assembled Quantum Materials and Grant No.~DMR-2011738. 
 We acknowledge computing resources from Columbia University’s
Shared Research Computing Facility project, which is supported by NIH Research
Facility Improvement Grant 1G20RR030893-01, and associated funds from the New
York State Empire State Development, Division of Science Technology and
Innovation (NYSTAR) Contract C090171, both awarded April 15, 2010.  The Flatiron
Institute is a division of the Simons Foundation.

\bibliography{anharm,prxrefs}
\end{document}